\documentclass[review]{elsarticle}

\journal{Renewable and Sustainable Energy Reviews}

\usepackage[top=1.25in, bottom=1.25in, left=1.25in, right=1.25in]{geometry}
\usepackage{csquotes}
\MakeOuterQuote{"}
\usepackage{amsmath, amssymb}
\usepackage{algorithm, algorithmic, verbatim, array, multirow, paralist, nicefrac, hyperref, setspace}
\usepackage{subcaption, graphicx, pgfplots, tikz, xcolor, setspace, adjustbox, float}
\usepackage[font=normalsize,labelfont=bf]{caption}
\pgfplotsset{compat=1.15}

\newcolumntype{C}[1]{>{\centering\let\newline\\\arraybackslash\hspace{0pt}}m{#1}}

\makeatletter
\def\ps@pprintTitle{%
   \let\@oddhead\@empty
   \let\@evenhead\@empty
   \def\@oddfoot{\reset@font\hfil\thepage\hfil}
   \let\@evenfoot\@oddfoot
}
\makeatother

\begin{document} 

\begin{frontmatter}
\title{An Integrated Optimization Framework for Multi-Component Predictive Analytics in Wind Farm Operations \& Maintenance}

\author[mymainaddress]{Bakir, I.}
\ead{i.bakir@rug.nl}

\author[mysecondaryaddress]{Yildirim, M. \corref{correspondingauthor}\fnref{myfootnote}}
\cortext[correspondingauthor]{Corresponding author}
\ead{murat@wayne.edu}
\fntext[myfootnote]{The first two authors contributed equally to this work.}

\author[mymainaddress]{Ursavas, E.}
\ead{e.ursavas@rug.nl}

\address[mymainaddress]{Department of Operations, Faculty of Economics and Business, \\ University of Groningen, Groningen, Netherlands}
\address[mysecondaryaddress]{Industrial and Systems Engineering, Wayne State University, Detroit, MI, USA}

\begin{abstract}
Recent years have seen an unprecedented growth in the use of sensor data to guide wind farm operations and maintenance. Emerging sensor-driven approaches typically focus on optimal maintenance procedures for single turbine systems, or model multiple turbines in wind farms as single component entities. In reality, turbines are composed of multiple components that dynamically interact throughout their lifetime. These interactions are central for realistic assessment and control of turbine failure risks. In this paper, an integrated framework that combines i) real-time degradation models used for predicting remaining life distribution of each component, with ii) mixed integer optimization models and solution algorithms used for identifying optimal wind farm maintenance and operations is proposed. Maintenance decisions identify optimal times to repair every component, which in turn, determine the failure risk of the turbines. More specifically, optimization models that characterize a turbine's failure time as the first time that one of its constituent components fail - a systems reliability concept called \emph{competing risk} is developed. The resulting turbine failures impact the optimization of wind farm operations and revenue. Extensive experiments conducted for multiple wind farms with 300 wind turbines - 1200 components - showcases the performance of the proposed framework over conventional methods.

\section*{Highlights}
\begin{itemize}
\item A condition-based maintenance and operations model is proposed for wind farms.

\item Component and turbine dependencies on failure risks and maintenance are modeled.

\item A tailored solution algorithm is proposed to ensure computational scalability.

\item A comprehensive experimental framework is developed via degradation and wind data.

\item The proposed approach provides significant improvements over benchmark models.
\end{itemize}
\end{abstract}

\begin{keyword}
Degradation models, multi-component reliability, condition based maintenance, wind farm operations and maintenance, large scale mixed integer optimization
\end{keyword}

\end{frontmatter}


\section{Introduction}\label{intro}
Maintenance scheduling is a fundamental component of wind farm operations with far reaching implications for equipment costs, market revenue, and maintenance crew logistics. In typical applications, maintenance costs constitute $25\%$ to $35\%$ percent of the running costs of wind farms \cite{Blanco2009,el2012operation,Sarker2016}. To mitigate this cost, wind farm operators are continuously looking into methods to improve the effectiveness of maintenance policies through use of real-time sensor data \cite{Wymore2015}. Condition monitoring (CM) systems identify indicators of component degradation within sensor data to develop accurate estimates for equipment condition. 
In traditional applications, CM systems alert the operators when the level of degradation in turbine components reach a certain level. These alerts are used to initiate immediate maintenance actions. Such policies rely solely on the current degradation state of the components. In reality, the complex relationships between component repair schedules, maintenance crew logistics and revenue opportunities, require a proactive organization of wind farm operations and maintenance (O\&M) \cite{Leite2018}. Prognostics based methods model the remaining life distribution of the components as they degrade, and provide the ability to plan maintenance and related activities before failure risks become eminent. 
This paper shows that when component-level prognostic predictions are integrated into wind farm operations and maintenance, the operators can sense early signs of degradation in every component, and jointly optimize O\&M activities by carefully considering the complex stochastic and economic inter-dependencies among multiple components, turbines and wind farms. 

Despite their importance, wind farm O\&M policies typically rely on ad-hoc decisions
in practice \cite{Leite2018, Erguido2017}.
Currently the typical form of maintenance policies for wind farm maintenance is a combination of strictly corrective policies that initiate maintenance actions upon the observation of failure, and time-based policies that perform maintenance at fixed time intervals \cite{Ding2011}. In wind farm O\&M literature, these policies are often augmented by the use of \emph{opportunistic maintenance}, which groups maintenance actions together to reduce crew visits, production losses, traveling and setup expenses \cite{Sarker2016}. Within opportunistic maintenance, scheduled corrective and preventive maintenance tasks are used as an opportunity to perform additional maintenance tasks on other components within a single turbine, or in multiple turbines within a farm. 

Conventional approaches to opportunistic maintenance in wind farms can be categorized under two approaches. The first approach uses optimization models to capture different aspects of operations and maintenance \cite{Abdollahzadeh2016,Besnard2009,Besnard2011}. The second approach \cite{Ding2012} provides structured policies to establish procedural maintenance decisions. These approaches establish multiple threshold values in terms of age or operational conditions that identify optimal times to initiate the first and the subsequent maintenance actions. A drawback of these conventional policies is that they do not use sensor data, and assume that the failure of identical components follow the same distribution. In reality, every component exhibit significant variation in terms of how they degrade and fail, due to many factors including manufacturing variations/defects, operational loading, and material imperfections. 

Recently, there has been a growing literature on using sensor data to infer accurate estimates for failure likelihood of turbines and their constituent components. An extensive survey of sensor-driven failure prediction models can be found in \cite{lei2018machinery}. These models rely on obtaining sensor-based indicators called \emph{degradation signals} that exhibit a strong correlation to the evolving degradation processes. \emph{Degradation models} develop stochastic formulations based on these signals to predict the remaining life distribution. Degradation modeling approaches for wind turbines are straightforward due to the relatively simple mechanical construction of wind turbine components \cite{hameed2009condition}. Integration of these life predictions into maintenance policies, however, still remains a fundamental challenge.

The earliest research efforts that incorporate sensor data to wind farm maintenance had the objective of  justifying the additional expense of condition monitoring equipment for individual components \cite{Nilsson2007,McMillan2008}. Several authors extended on these papers to develop methods for sensor-driven maintenance and operations. Majority of the work in the area focuses on detailed modeling of sensor-driven maintenance for single turbine systems. \cite{byon2010optimal} considers the sensor-driven maintenance of gearboxes under imperfect maintenance actions. This work is extended in \cite{byon2010season} by enabling the maintenance decisions to adapt to season-dependent and time-varying weather conditions. These studies demonstrate that considering seasonality when making maintenance decisions helps in decreasing failure rates, as well as reducing O\&M costs. \cite{Besnard2010, byon2010optimal, byon2010season} all consider a single component type, respectively blades and gearboxes. While these policies provide significant developments in the field and consider important factors that impact degradation, they do not necessarily capture complex degradation-based and economic interactions between multiple turbines and components. As suggested by \cite{Yildirim2017a}, sensor-driven maintenance policies that focus on a single turbine, may not scale well to complex wind farm setting where the component interactions are significant. In fact, in certain cases sensor-driven single turbine policies may perform worse than simpler periodic maintenance policies that accurately model the wind farm as a whole \cite{Yildirim2017a}.

The works that consider sensor-driven policies as well as opportunistic maintenance are scarce. One of the first studies to consider the actual condition of multiple components when making maintenance decisions is \cite{Tian2011}. In this study, a maintenance policy defined by two thresholds is proposed, where the first threshold initiates preventive maintenance, and the second - lower - threshold is used for opportunistic preventive maintenance. In a recent paper, \cite{Zhou2019} proposed a new method to identify optimal two threshold policies using a detailed model of offshore maintenance cost parameters. A key assumption in these papers is that the thresholds are static. While threshold-optimization policies provide an important and practical approach, they do not necessarily reflect that these thresholds exhibit significant variation across different time periods and turbines due to multitude of factors such as crew management, market price, and wind uncertainty. 
In the recent work of \cite{Perez2015} a wind farm with 100 turbines is considered on a component level. However, the developed maintenance policy only considers opportunistic maintenance of multiple components within the same wind turbine, neglecting possible dependencies between turbines. 
The dependencies between turbines are considered in \cite{Yildirim2017a} that proposes an optimization framework for the O\&M scheduling for wind farms. Although this work models dependencies between wind turbines, the dependencies between components within wind turbines are neglected.

Literature on opportunistic and sensor-driven maintenance approaches clearly demonstrate that these two approaches provide significant benefits when applied in isolation. The benefits of their integration, however, can be studied in more detail: 
The studies considering opportunistic maintenance are mostly time-based, and the studies that consider sensor-driven approaches are predominantly aimed at component- or, single-turbine-level. The studies that consider opportunistic and sensor-driven maintenance together on a wind farm level either neglected the inter-dependencies between turbines \cite{Perez2015}, or between components \cite{Yildirim2017a}. To the best of the authors' knowledge, there is no optimization framework that develops opportunistic and sensor-driven (\emph{i.e. prognostics-based}) maintenance for wind farms that accurately model the complex interactions between failures and maintenance actions at a component level. This forms the main contribution of the framework.

To bridge this gap, this work proposes a unified framework that integrates component-level prognostics into large-scale mixed integer optimization models used for identifying optimal wind farm maintenance and operations decisions. The proposed model considers component-level maintenance actions to capture the inter-dependencies among components within a turbine, as well as components across turbines in multiple wind farms. A unique aspect of this approach is the integration of opportunistic maintenance across three levels: components, turbines, and multiple wind farms.  Main contributions of the paper are listed below:
\begin{enumerate}
\item A novel wind farm operations and maintenance framework that adapts to real-time sensor data to capture dynamic interactions among turbine components, turbines and wind farms is developed. The proposed framework offers a significant shift from existing sensor driven policies that i) focus on single wind turbine systems with multiple components, or ii) model maintenance of multiple wind farms without considering component-level interactions.
\item Component degradation models are integrated within a novel mixed integer decision optimization formulation used for identifying optimal wind farm operations and maintenance. This approach models dynamically evolving turbine failure risks as a function of i) real-time predictions on component remaining life distributions, and ii) component-level maintenance decisions. Component maintenance decisions and the resulting turbine failure risks are tightly integrated into O\&M of multiple wind farms, and the routing for the maintenance crew. 
\item Modeling enhancements and algorithmic approaches to solve large instances of the resulting decision optimization model is proposed. More specifically, a linearization that enforces a one-to-one mapping between component-level maintenance decisions and the state of the turbine system is developed. The model is decomposed into a two-stage form, where the master level problem consists of wind farm- and turbine-level decisions, and the subproblem identifies decisions for the constituent components. The resulting reformulation enables a master-slave type solution algorithm based on Benders' and integer cuts.
\end{enumerate}
A comprehensive experimental platform that models the impact of different maintenance and operations decisions is developed. Degradation processes in turbine components are  emulated by using a database of vibration based degradation data collected from rotating machinery. Wind is modeled by using the database from KNMI North Sea Wind Atlas project \cite{KNMI2014}. The complex degradation based and economic interactions between multiple components, wind turbines, and wind farms are simulated. Extensive set of experiments suggest that the proposed policy provides significant advantages over existing approaches, in terms of cost, reliability and the efficient use of equipment lifetime.

The remainder of the paper includes the following: Section \ref{sec:methodology} introduces the method for the unified framework, which covers degradation modeling for multi-component turbine systems, and its integration into a novel large scale optimization model. Section \ref{sec:algorithm} introduces the solution algorithm and highlights its computational benefits. Section \ref{sec:experiments} presents the experimental results for benchmark and robustness studies. Finally, the conclusion is provided in Section \ref{sec:conclusion}. 

\section{Method} \label{sec:methodology}

This section formally introduces the integrated wind farm operations and maintenance scheduling framework that is composed of two main modules: i) sensor-driven multi-component degradation models for quantifying failure risks, and ii) sensor-driven adaptive optimization for multi-component opportunistic maintenance and operations. These modules are tightly integrated, i.e. sensor-driven component remaining life predictions continuously update the operations and maintenance scheduling for the wind farm. 

The proposed framework assumes that condition monitoring technology exists for wind turbine components to enable the predictions of component-level remaining life distributions. Condition monitoring, diagnostics and prognostics in wind turbines has a rich and growing literature. For more details on this literature, we refer the reader to \cite{kandukuri2016review,leite2018prognostic,lau2012review,salameh2018gearbox,wang2019vibration,hameed2009condition}. 

To elucidate the framework in detail, this section demonstrates how component specific low-level sensor observations are used to update a series of modules that drive high-level decisions in wind farm operations and maintenance.

\subsection{Sensor-Driven Multi-Component Degradation Models for Quantifying Failure Risks}

The first task is to use real-time sensor information to derive accurate predictions on when the turbines and their constituent components are likely to fail. This requires a two step procedure. The first step is to use the component specific sensor data to predict remaining life distribution of all the components in a turbine. A mapping between component remaining life and the failure probabilities of the turbine is then derived. This degradation framework enables the quantification of failure risks both at a component and a turbine level.

\subsubsection{Component Remaining Life Prediction} 
Turbine components degrade over time due to use and aging. Degradation models characterize this degradation processes in detail to improve failure risk predictions. This paper uses a generalized parametric degradation model for component $k$ in turbine $i$ in the form of the function: ${D}_{i,k}(t) = \phi_{i,k}(t; {\kappa}, {\theta}_{i,k}) + \epsilon_{i,k}(t; {\sigma})$. In this formulation, $\phi_{i,k}(t;{\kappa}, {\theta}_{i,k})$ defines the underlying base degradation function for component $k$ of turbine $i$, and $\epsilon_{i,k}(t; {\sigma})$ models the uncertainty due to degradation and measurement errors, with an associated variance parameter ${\sigma}$. The parameters $\kappa$ and $\theta_{i,k}$ denote the deterministic and stochastic degradation parameters, respectively. 

It is assumed that engineering knowledge and historical data yield an initial estimate for the distribution of the stochastic degradation parameter $\theta_{i,k}$, denoted by $\pi(\theta_{i,k} )$. Further assumption is that the operators observe a set of degradation signals 
that enable the update of the distribution of the stochastic parameter $\theta_{i,k}$ to its posterior distribution counterpart $\upsilon( {\theta}_{i,k})$ using Bayesian update. 
Given the updates on the degradation parameters, remaining life distribution of component $k$ in turbine $i$ can be computed as the first time that the degradation signal crosses a predefined failure threshold $\Lambda_{i,k}$. 
The remaining life of component $k$ in turbine $i$, namely $\tau_{i,k}$ can be evaluated as $P(\tau_{i,k}  = t) = P( t = min(s\geq 0 | D_{i,k}(s|\upsilon( {\theta}_{i,k})) \geq  \Lambda_{i,k} )$. See \cite{yildirim1} for more details on this class of degradation models. 

These remaining life predictions drive two submodules that are used by the operations and maintenance optimization model: i) dynamic component maintenance cost functions, and ii) turbine remaining life predictions.

\subsubsection{Dynamic Component Maintenance Cost Function} \label{costfunction}
Given the updated remaining life distributions of component $k$, the expected cost of conducting maintenance on the component at time $t$ is derived, as follows \cite{yildirim1,elwany2008sensor}: \begin{align} \label{eq:Cold}
C^{i,k}_{t^o,t} = \frac{ c^p_{i,k} P( \tau_{i,k} > t) + c^f_{i,k} P( \tau_{i,k} \leq t) }{ \int_0^{t} P( \tau_{i,k} > z) dz + t_o },
\end{align}
where $c_{i,k}^p$ and $c_{i,k}^f$ are the costs for preventive maintenance, and unexpected failure for component $k$, respectively. The function translates the remaining life distribution of component $k$ into a degradation-based function of expected cost over time. 

\subsubsection{Turbine Remaining Life Prediction} 
Predictions on the component remaining life distribution are used to derive failure likelihood of turbines as well. Assume that $\mathcal{K}_i$ is the set of critical components in turbine $i$. The failure time of turbine $i$, $\tau^i$ can be defined as the first time that one of its critical components fail:
\begin{align*}\label{eq:ProbRi}
&P( \tau^i > t) = \int P\left( \max\limits_{k \in \mathcal{K}_i} \sup\limits_{t_o \leq s\leq t_o + t} \varphi(s,\theta_{i,k})  >0 \right) 
\prod_{k \in \mathcal{K}_i} \upsilon({\theta}_{i,k}) d{\theta}_{i,k}
\end{align*}
where $\varphi(s,\theta_{i,k}) = {D}_{i,k}(s| {\theta}_{i,k}) - \Lambda_{i,k}$, and $t_o$ is the age of the component. This failure definition uses a well known failure modeling concept called competing risk \cite{crowder2001classical, lindqvist2008modeling, peng2010reliability}. 

\subsubsection{Impact of Component Maintenance on Turbine Failure Risks}\label{Sec:turbdegradation}

Turbine failure risks are highly dependent on component maintenance schedules. There are two main assumptions to capture the impact of maintenance on turbine failure risks: i) a component becomes as good as new after maintenance, ii) a maintained component does not fail again within the planning horizon. Second assumption is reasonable so long as the planning horizon doesn't exceed an annual span. In fact, it is trivial to augment the proposed model to eliminate the second assumption, which is chosen to be enforced here for notational convenience.

The maintenance time for component $k$ in turbine $i$ is denoted as $\chi_{i,k}$. This is an integer value: assume that the maintenances may take place at the beginning of each period. The probability that a turbine survives until time $t$ is:
\begin{align*}
 P\left( \tau^i >t \right)  = P\left( \tau_{i,k} >min(t,\chi_{i,k}) \quad \forall k \in \mathcal{K}_i \right)
\end{align*}
A key observation from this equation is that the probability of failure during $(t,t+1]$, directly depends on whether a maintenance was conducted at or before $t$, i.e. $\varphi_{i,k} \leq t$. This means that the set of possible maintenance scenarios at each time period $(t,t+1]$, has a cardinality of $2^{|\mathcal{K}_i|}$; i.e. each component has two states: maintained or not maintained. It is not computationally demanding to precompute the turbine failure probabilities of these maintenance scenarios for each time period and turbine. This is an important point for integrating the model to optimization; i.e. only $2^{|\mathcal{K}_i|}$ events for time $t$ are considered, as opposed to the entire history with $2^{t \cdot |\mathcal{K}_i|}$ unique events. 
\begin{figure}
\centering
\includegraphics[width=0.8\textwidth]{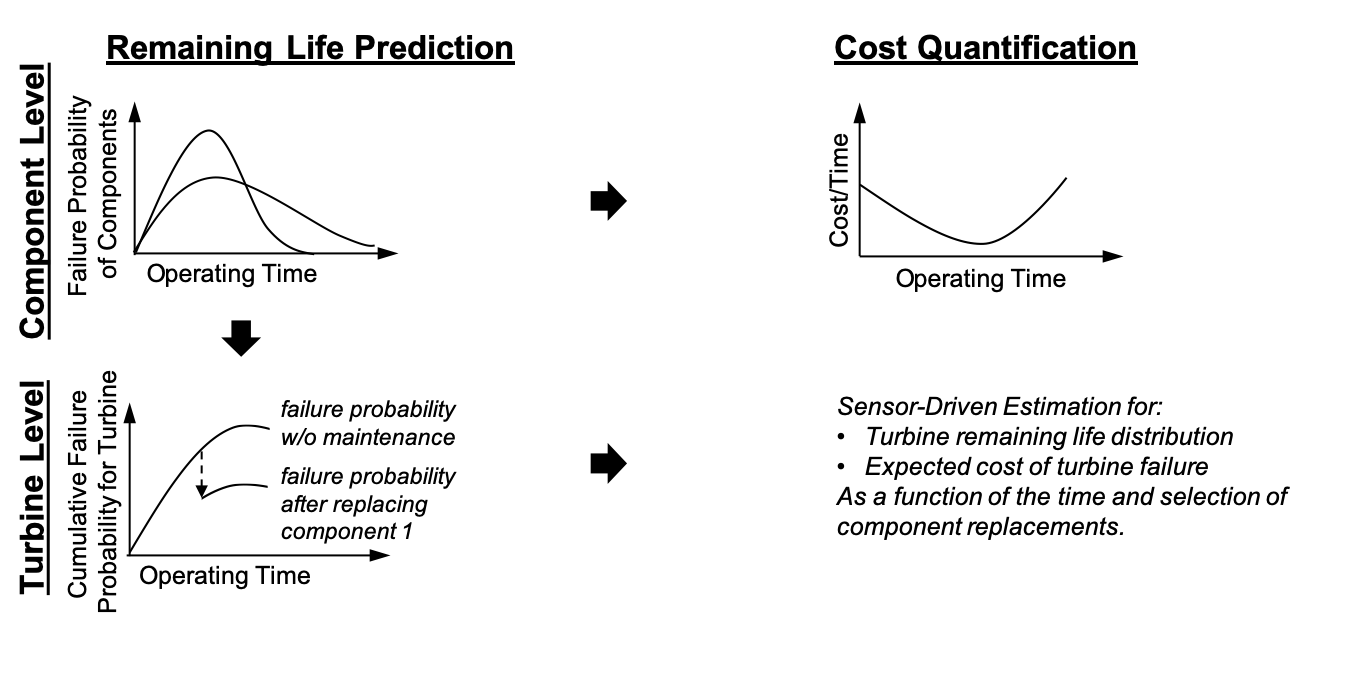}
\caption{Integrating {Component \&} {Turbine-Level} Risks.
}
\label{fig:gencvar}
\end{figure}
\subsection{Sensor-Driven Adaptive Optimization for Multi-Component Opportunistic Maintenance and Operations}
Given the predictive degradation models, expectations on failure likelihoods and associated maintenance costs, the next challenge is to develop a fully-adaptive and comprehensive, multi-windfarm optimization model for operations and maintenance scheduling. The proposed Multi-Component Condition Based Opportunistic Maintenance and Operations Model (MC-CBOM) tightly integrates the dynamic maintenance cost functions and the predicted likelihoods of failure for every component and turbine. Figure \ref{fig:gencvar} highlights the information flow across component and turbine-level analytics that feed into the optimization models. Information exchange occurs in two levels. At a component level, the dynamic maintenance cost function \eqref{eq:Cold} is discretized for each time period and integrated into the objective function of the optimization model. Turbine failure probabilities are reconstructed within the optimization model as a function of maintenance decisions and component failure probabilities. More specifically, variables associated with maintenance decisions and turbine maintenance scenarios (\emph{as introduced in Section \ref{Sec:turbdegradation}}) enable the optimization model to capture the impact of maintenance decisions on turbine failure probability. The coupling between turbine maintenances and operations is also explicitly captured within the framework.

More specifically, the proposed model considers multiple layers of dependencies across components and turbines. In what follows, these dependencies are motivated with practical examples:

\emph{Component-level dependencies:} When a maintenance team visits a turbine to fix a certain component, it may make sense to fix another component that may also be highly degraded. A simplified example could be that a blade repair order may also include a fix to another blade or a yaw control system that may also be highly degraded. This would be particularly important when the team has a significant setup cost associated with setting up the crane to access the turbine components.

\emph{Component-to-turbine dependencies:} Turbine failure is continuously characterized as a function of its constituent components. A maintenance action on a component, would have a direct impact on component failure probability, and an indirect impact on the failure likelihood of the turbine as whole. Both of these aspects are explicitly modeled in the MC-CBOM.

\emph{Turbine-level dependencies:} Given maintenance crew routing requirements, only a subset of turbines that are located in the same wind farm location $l$ can be maintained simultaneously. It would therefore be important to understand the cumulative impact of failure risks from multiple turbines before a crew visits a certain location within a wind farm.

Before formally developing the model, the notation is introduced. Let $\mathcal{T}$, $\mathcal{L}$, $\mathcal{G}$, and $\mathcal{K}$ denote the sets of time periods, wind farm locations, wind turbines, and turbine components, respectively. Set $\mathcal{G}^l$ represents the wind turbines in wind farm $l$, and its two subsets $\mathcal{G}^l_o$ and $\mathcal{G}^l_f$ denote the operational and failed turbines at the start of the planning horizon. A turbine is classified as operational if all of its components are operational, and failed if at least one of its components is in a failed state. For each turbine $i$ within wind farm $l$, the set of components is also partitioned into subsets $\mathcal{K}^{l,i}_o$ and $\mathcal{K}^{l,i}_f$, representing operational and failed components of turbine $i$ at the beginning of the planning horizon, respectively. 

To denote the preventive maintenance decisions, the binary decision variable $z$ is used, which is defined only over the set of currently operational components. Variable $z_t^{l,i,k}$ will assume the value 1 if preventive maintenance on currently operational component $k$ of turbine $i$ in wind farm location $l$ is initiated at period $t$ (and 0 otherwise). The preventive maintenance decisions incur a dynamic maintenance cost, as given in equation \eqref{eq:Cold}. 
Corrective maintenance actions are represented with binary variable $\nu$, which is defined only over the set of components that are currently at a failed state. The variable $\nu_t^{l,i,k}$ will assume the value 1 if corrective maintenance on component $k$ of turbine $i$ in wind farm $l$ is initiated at period $t$ (and 0 otherwise).

Visits of the maintenance crew to wind farm locations and wind turbines are denoted with binary decision variables $x$ and $\alpha$, respectively. If wind farm location $l$ is visited by the maintenance crew at period $t$, then $x_t^l$ will be 1; and if that visit included turbine $i$, then $\alpha_t^{l,i}$ will also be 1. The crew deployment costs for wind farm and turbine visits are denoted with $V^{v,l}_{F,t}$ and $V^{v,l,i}_{T,t}$, respectively. Energy generation is represented with a continuous decision variable $y$, where $y_t^{l,i}$ denotes the energy (in MWh) generated by wind turbine $i$ in period $t \in \mathcal{T}$. There is a revenue $\pi_t$ associated with each generated MWh of electricity.

The remainder of this section details various components, i.e., objective function and constraints, of the MC-CBOM.

\subsubsection{Objective Function}
The objective of the model is to maximize the total profit from all wind farm locations throughout the planning horizon. The objective function \eqref{full_obj} calculates the net profit by subtracting crew deployment, turbine maintenance, and expected turbine failure costs from  the operational revenue. 

\vspace{-10pt}
\begin{equation}
\begin{aligned}
\max
&  \sum_{\substack{l \in \mathcal{L},  i \in \mathcal{G}^l,  t \in \mathcal{T}}} \left( \pi_t y_t^{l,i} - V_{T,t}^{v,l} \alpha_t^{l,i} \right) - \sum_{\substack{l \in \mathcal{L}, t \in \mathcal{T}}} V_{F,t}^{v,l} x_t^l \\
& \hspace{2mm} 
- \sum_{\substack{l \in \mathcal{L},  i \in \mathcal{G}^l_o, \\ k \in \mathcal{K}_o^{l,i},  t \in \mathcal{T}}} C_{t_k^o,t}^{l,i,k} z_t^{l,i,k} - \sum_{\substack{l \in \mathcal{L},  i \in \mathcal{G}^l_o, \\ t \in \mathcal{T}}} C^{f,l,i} \rho_t^{l,i} \label{full_obj}
\end{aligned}
\end{equation}
\vspace{-10pt}

Note that every time the model is constructed, the dynamic maintenance costs, denoted with $C_{t_k^o,t}^{l,i,k}$, are computed using the remaining life predictions of operational wind turbine components and updated with recent sensor readings. These costs are crucial in making preventive maintenance decisions. 

The last term of the objective function evaluates the costs associated with turbine failure based on actual failure probabilities of the components within the turbine. Here, $\rho_t^{l,i}$ represents the failure probability of turbine $i$ of wind farm $l$ in time period $t$. This variable is multiplied with a failure cost $C^{f,l,i}$ to obtain the expected failure cost.

\subsubsection{Component Maintenance Coordination and Coupling Constraints}
Constraints \eqref{C5} mandate that a component undergoes preventive maintenance before its reliability drops below a prespecified threshold $\bar{\zeta}$. To enforce this requirement, a time limit $\zeta^{l,i,k} := \min \{t \in \mathcal{T} : P(R^{l,i,k}_{t_i^o} > t) < \bar{\zeta} \}$ is computed for all components; and constraints \eqref{C5} ensure that a preventive maintenance is initiated on or before this time limit. Constraints \eqref{C6} enforce that at most one corrective maintenance is conducted on each failed component during the planning horizon.

\vspace{-30pt}
\begin{align}
& \sum_{t=1}^{\zeta^{l,i,k}} z_t^{l,i,k} = 1, \qquad \forall l \in \mathcal{L}, \forall i \in \mathcal{G}^l, \forall k \in \mathcal{K}_o^{l,i} \label{C5} \\
& \sum_{t \in \mathcal{T}} \nu_t^{l,i,k} \leq 1, \qquad \forall l \in \mathcal{L}, \forall i \in \mathcal{G}^l, \forall k \in \mathcal{K}_f^{l,i} \label{C6}
\end{align}
\vspace{-20pt}

Turbine maintenance is denoted with binary variable $\alpha$, where $\alpha_t^{l,i} = 1$ indicates that a maintenance activity (preventive or corrective) is being conducted on turbine $i$ in period $t$. Constraints \eqref{C7} and \eqref{C8} ensure the coupling of turbine visit and component maintenance decisions. This way, a turbine visit is scheduled if any of the turbine's components undergoes preventive or corrective maintenance. Constraints \eqref{C10} couple the wind farm and turbine visit variables by enforcing that a crew visit is planned for a wind farm location whenever a turbine in that location is to be visited by the maintenance crew.

\vspace{-30pt}
\begin{align}
& z_t^{l,i,k} \leq \alpha_t^{l,i}, \qquad \forall l \in \mathcal{L}, \forall i \in \mathcal{G}^l, \forall k \in \mathcal{K}_o^{l,i}, \forall t \in \mathcal{T} \label{C7} \\
& \nu_t^{l,i,k} \leq \alpha_t^{l,i}, \qquad \forall l \in \mathcal{L}, \forall i \in \mathcal{G}^l, \forall k \in \mathcal{K}_f^{l,i}, \forall t \in \mathcal{T} \label{C8} \\
& \alpha_t^{l,i} \leq x_t^l, \qquad \quad \forall l \in \mathcal{L}, \forall i \in \mathcal{G}^l, \forall t \in \mathcal{T} \label{C10}
\end{align}
\vspace{-30pt}

\subsubsection{Maintenance Crew Logistics} Constraints \eqref{C11} ensure that the same crew cannot visit multiple locations at the same time period. Constraints \eqref{C12} limit the crew visit to location $l$, when the conditions do not allow a visit.

\vspace{-30pt}
\begin{align}
& \sum_{l \in \mathcal{L}} x_t^l \leq 1, \qquad \forall t \in \mathcal{T}, \label{C11} \\
& x_t^l = 0, \qquad \qquad \forall l \in \mathcal{L}, \forall t \in \bar{\mathcal{T}}^l, \label{C12}
\end{align} 
\vspace{-25pt}

where $\bar{\mathcal{T}}^l$ denotes the set of time periods when a maintenance crew visit to location $l$ is not possible due to limiting conditions, such as weather-related or organizational restrictions. Constraints \eqref{C13} enforce the required travel time for traveling between location $l$ and $l'$. 

\vspace{-30pt}
\begin{align}
x_t^l + x_{t'}^{l'} \leq 1, \quad & \forall l \in \mathcal{L}, \forall l' \in \mathcal{L} \setminus \{l\}, \forall t \in \{1, \ldots, T - \theta_{l,l'}\}, \nonumber \\
& \forall t' \in \{t, \ldots, t + \theta_{l,l'}\} \label{C13}
\end{align}
\vspace{-25pt}

where $\theta_{l,l'}$ denotes the required time to travel from location $l$ to $l'$.

\subsubsection{Maintenance Capacity}
Constraints \eqref{C9} limit the number of turbines on which maintenance is being conducted during time period $t$ to the maintenance capacity at that time period, which is denoted with $M_t$.

\vspace{-20pt}
\begin{equation}
\sum_{l \in \mathcal{L}} \sum_{i \in \mathcal{G}^l} \alpha_t^{l,i} \leq M_t, \qquad \forall t \in \mathcal{T} \label{C9}
\end{equation}
\vspace{-10pt}

\subsubsection{Energy Generation}
These constraints couple the energy generation decision variables, $y$, with component maintenance decision variables, $z$ and $\nu$. Constraints \eqref{C14} allow operational turbine $i$ in location $l$ to produce up to wind-induced production capacity at time $t$, namely $p_t^{l,i}$, unless the turbine has an ongoing maintenance.

\vspace{-20pt}
\begin{equation}
y_t^{l,i} \leq p_t^{l,i} (1 - \alpha_t^{l,i}), \qquad \forall l \in \mathcal{L}, \forall i \in \mathcal{G}^l, \forall t \in \mathcal{T} \label{C14}
\end{equation}
\vspace{-20pt}

Constraints \eqref{C15} enforce that a failed wind turbine cannot produce electricity until it undergoes corrective maintenance. 
Once a turbine that started at a failed state is correctively maintained, it can produce up to $p_t^{l,i}$. Note that this decision considers the impact of turbine outage on production loss.

\vspace{-20pt}
\begin{equation}
y_t^{l,i} \leq p_t^{l,i} \sum_{\tau=1}^{t-1} \alpha_{\tau}^{l,i},\qquad \forall l \in \mathcal{L}, \forall i \in \mathcal{G}^l_f, \forall t \in \mathcal{T} \label{C15}
\end{equation}
\vspace{-10pt}

\subsubsection{Turbine Failure Probability} The following constraints determine failure probabilities for each turbine $i$ based on the condition and maintenance status of each of its components. The binary variable $u$ denotes whether or not a component underwent preventive or corrective maintenance before time period $t$; $u^{l,i,k}_t = 1$ if component $k$ of turbine $i$ in wind farm $l$ has been maintained before time period $t$. Constraints \eqref{C1} and \eqref{C2} establish this relationship between these variables and the maintenance scheduling variables $z$ and $\nu$.

\vspace{-30pt}
\begin{align}
& \sum_{\tau=1}^{t-1} z_{\tau}^{l,i,k} = u_t^{l,i,k}, \quad \forall l \in \mathcal{L}, \forall i \in \mathcal{G}, \forall k \in \mathcal{K}_o^i, \forall t \in \mathcal{T} \label{C1} \\
& \sum_{\tau = 1}^{t-1} \nu_{\tau}^{l,i,k} = u_t^{l,i,k}, \quad \forall l \in \mathcal{L}, \forall i \in \mathcal{G}, \forall k \in \mathcal{K}_f^i, \forall t \in \mathcal{T} \label{C2}
\end{align}
\vspace{-20pt}

Failure probability of a turbine depends on the age and maintenance status of each of its components. Let $\mathcal{H}^{l,i}$ define a set which contains all possible maintenance scenarios for each turbine $i$ in wind farm $l$. If a turbine consists of $n$ components, $2^n$ maintenance scenarios would be possible. For example, when a turbine consists of 2 components, the 4 possible maintenance scenarios are {\em(i)} no maintenance is conducted on any of the components, {\em(ii)} maintenance is conducted only on component 1, {\em(iii)} maintenance is conducted only on component 2, and {\em(iv)} both components are maintained. Binary variable $\eta$ is used to denote the maintenance status of a turbine, based on the maintenance status of each of its components. More specifically, $\eta^{l,i,h}_t = 1$ indicates that maintenance scenario $h$ is realized on turbine $i$ of wind farm $l$ in time period $t$. Constraints \eqref{C3} establish this relationship.

\vspace{-30pt}
\begin{align}
\eta_t^{l,i,h} \geq \sum_{k \in \mathbb{K}^{l,i,h}} & u_t^{l,i,k} - \sum_{k \in \mathbb{F}^{l,i,h}} u_t^{l,i,k} - | \mathbb{K}^{l,i,h} | + 1, \nonumber \\
& \forall l \in \mathcal{L}, \forall i \in \mathcal{G}^l, \forall h \in \mathcal{H}^{l,i}, \forall t \in \mathcal{T} \label{C3}
\end{align}
\vspace{-20pt}

where $\mathbb{K}^{l,i,h}$ ($\mathbb{F}^{l,i,h}$) denotes the set of components of turbine $i$ that have (not) been maintained under maintenance scenario $h$. Recall that the continuous variable $\rho_t^{l,i}$ denotes the failure probability of turbine $i$ of wind farm $l$ in period $t$. Constraints \eqref{C4} enforce the relationship between this variable and the maintenance scenario variable $\eta$, and ensure that if maintenance scenario $h$ is realized, the failure probability will take the corresponding value, $\phi^{l,i,h}_t$.

\vspace{-20pt}
\begin{equation}
\rho_t^{l,i} \geq \phi_t^{l,i,h} \eta_t^{l,i,h}, \quad \forall l \in \mathcal{L}, \forall i \in \mathcal{G}^l, \forall h \in \mathcal{H}^{l,i}, \forall t \in \mathcal{T} \label{C4}
\end{equation}
\vspace{-20pt}

\subsubsection{Decision Variable Limits} In the formulation, 
component maintenance variables $\mathbf{z}$ and $\boldsymbol{\nu}$, turbine visit variable $\boldsymbol{\alpha}$, maintenance crew visit variable $\mathbf{x}$, maintenance scenario variable $\boldsymbol{\eta}$, and cumulative maintenance variable $\mathbf{u}$ are binary. On the other hand, failure probability variable $\boldsymbol{\rho}$ and energy generation variable $\mathbf{y}$ are continuous. $\mathcal{Y}$ is used to indicate the feasible region for the variable limits:

\vspace{-30pt}
\begin{align}\label{C_varlims}
\mathbf{z}, \boldsymbol{\nu}, \boldsymbol{\alpha}, \mathbf{x}, \boldsymbol{\eta}, \mathbf{u}, \boldsymbol{\rho}, \mathbf{y} \in \mathcal{Y}.
\end{align}
\vspace{-20pt}

\section{Reformulation and Solution Algorithm} \label{sec:algorithm}

The proposed mathematical model is computationally complex and not all instances can be solved by a standard solver. Therefore the structural properties of the model is studied and a tailored technique that enables solving instances of realistic sizes to optimality is developed.

The solution method builds on a novel decomposition of the optimization model into a two-stage form. In the first stage the wind farm location- and turbine-level decisions are determined. At the second stage the decisions of the related wind turbine components are solved. This decomposition form is used to develop a  master-slave type solution algorithm based on Benders' and integer cuts - enabling a scalable integration of the proposed framework for wind farms.

A restricted master problem is formulated by incorporating 
decisions related to the wind farm location and the turbine levels, as well as an additional auxiliary variable $\Psi$ to represent a lower bound of the optimal value from the subproblem, which includes decisions related to turbine components. The subproblem is solved as a mixed integer programming model, and as a linear relaxation at different time steps of the algorithm. Solutions from the first and second type of subproblems are used to generate integer cuts, and Benders' cuts, respectively. These cuts are iteratively incorporated into to the restricted master problem to obtain an optimal solution in a finite number of iterations.

The restricted master problem is formulated as follows:

\vspace{-30pt}
\begin{align}
\max_{\boldsymbol{\alpha}, \mathbf{x}, \mathbf{y}} & \sum_{\substack{l \in \mathcal{L},  i \in \mathcal{G},  t \in \mathcal{T}}} \left( \pi_t y_t^{l,i} - V_{T,t}^{v,l} \alpha_t^{l,i} \right) - \sum_{\substack{l \in \mathcal{L},  t \in \mathcal{T}}} V_{F,t}^{v,l} x_t^l - \Psi \label{master_obj} \\
\text{s.t. } &  \eqref{C10}, \eqref{C11}, \eqref{C12}, \eqref{C13}, \eqref{C9}, \eqref{C14}, \eqref{C15}
\nonumber \\
& \Psi \geq \boldsymbol{\lambda}^T \boldsymbol{\alpha}, \qquad \forall \boldsymbol{\lambda} \in \Lambda^P \label{opt_cut} \\
& 0 \geq \boldsymbol{\lambda}^T \boldsymbol{\alpha}, \qquad \forall \boldsymbol{\lambda} \in \Lambda^R \label{feas_cut}
\end{align}
\vspace{-25pt}

The objective function \eqref{master_obj} maximizes the operational profit by explicitly accounting for the operational revenue and wind farm location- and turbine-level costs, while considering the auxiliary variable, $\Psi$, as a lower bound for all other costs. Note that these other costs, which are related to component-level decisions, are explicitly accounted for in the subproblem. The master problem includes constraints \eqref{C10}-\eqref{C15}, which dictate the operational requirements and relationships among the wind farm location- and turbine-level decisions. Constraints \eqref{opt_cut} collectively denote the Benders' optimality cuts and the integer cuts. Constraints \eqref{feas_cut} denote the Benders' feasibility cuts. The limits for the $\mathbf{x}, \mathbf{y}, \boldsymbol{\alpha}$ variables are consistent with those given in \eqref{C_varlims}.

When the master problem is solved, the values for the $\boldsymbol{\alpha}$ variables are used in generating the subproblem, which solves for the second stage variables related to wind turbine components. The formulation for the subproblem is given below: 

\vspace{-30pt}
\begin{align}
\min_{\mathbf{z}, \boldsymbol{\nu}, \mathbf{u}, \boldsymbol{\eta}, \boldsymbol{\rho}} & \sum_{\substack{l \in \mathcal{L},  i \in \mathcal{G}_o, \\ k \in \mathcal{K}_o^i,  t \in \mathcal{T}}} C_{t_k^o,t}^{l,i,k} z_t^{l,i,k} + \sum_{\substack{l \in \mathcal{L},  i \in \mathcal{G}_o, \\ t \in \mathcal{T}}} C^{f,l,i} \rho_t^{l,i} \label{sub_obj} \\
\text{s.t. } & \eqref{C5}, \eqref{C6}, \eqref{C1}, \eqref{C2}, \eqref{C3}, \eqref{C4}
\nonumber \\
& z_t^{l,i,k} \leq \bar{\alpha}_t^{l,i}, \quad \forall l \in \mathcal{L}, \forall i \in \mathcal{G}^l, \forall k \in \mathcal{K}_o^i, \forall t \in \mathcal{T} \label{sub_C7} \\
& \nu_t^{l,i,k} \leq \bar{\alpha}_t^{l,i}, \quad \forall l \in \mathcal{L}, \forall i \in \mathcal{G}^l, \forall k \in \mathcal{K}_f^i, \forall t \in \mathcal{T} \label{sub_C8}
\end{align}
\vspace{-25pt}

The objective function \eqref{sub_obj} minimizes the costs related to the second stage variables. The subproblem includes constraints \eqref{C5}, \eqref{C6}, \eqref{C1}-\eqref{C4} of the original model, which ensure maintenance coordination and manage the relationships between maintenance scenarios and failure probabilities. Constraints \eqref{sub_C7} and \eqref{sub_C8} enforce that the second stage decisions are consistent with the first stage decision variables, whose values are denoted with the $\bar{\boldsymbol{\alpha}}$ vector. It is important to note that the subproblem formulation is separable in turbines, which means that a separate model can be built and solved for each turbine $i \in \mathcal{G}^l$ in each location $l \in \mathcal{L}$. This property allows solving a number of small optimization problems, rather than solving one big problem, which enables parallel implementation and further scalability.

If a solution to the subproblem can be found, then the values of the dual variables corresponding to this solution are computed. This dual vector is then added to the set of optimality cut coefficients, $\Lambda^P$. These cuts will be later augmented by integer cuts as well. If, however, the subproblem has no feasible solutions, this means that the dual of the subproblem is unbounded, and thus an extreme ray is found and then added to the set of feasibility cut coefficients, $\Lambda^R$.

\begin{figure}[t!]
\centering
\includegraphics[width=\textwidth]{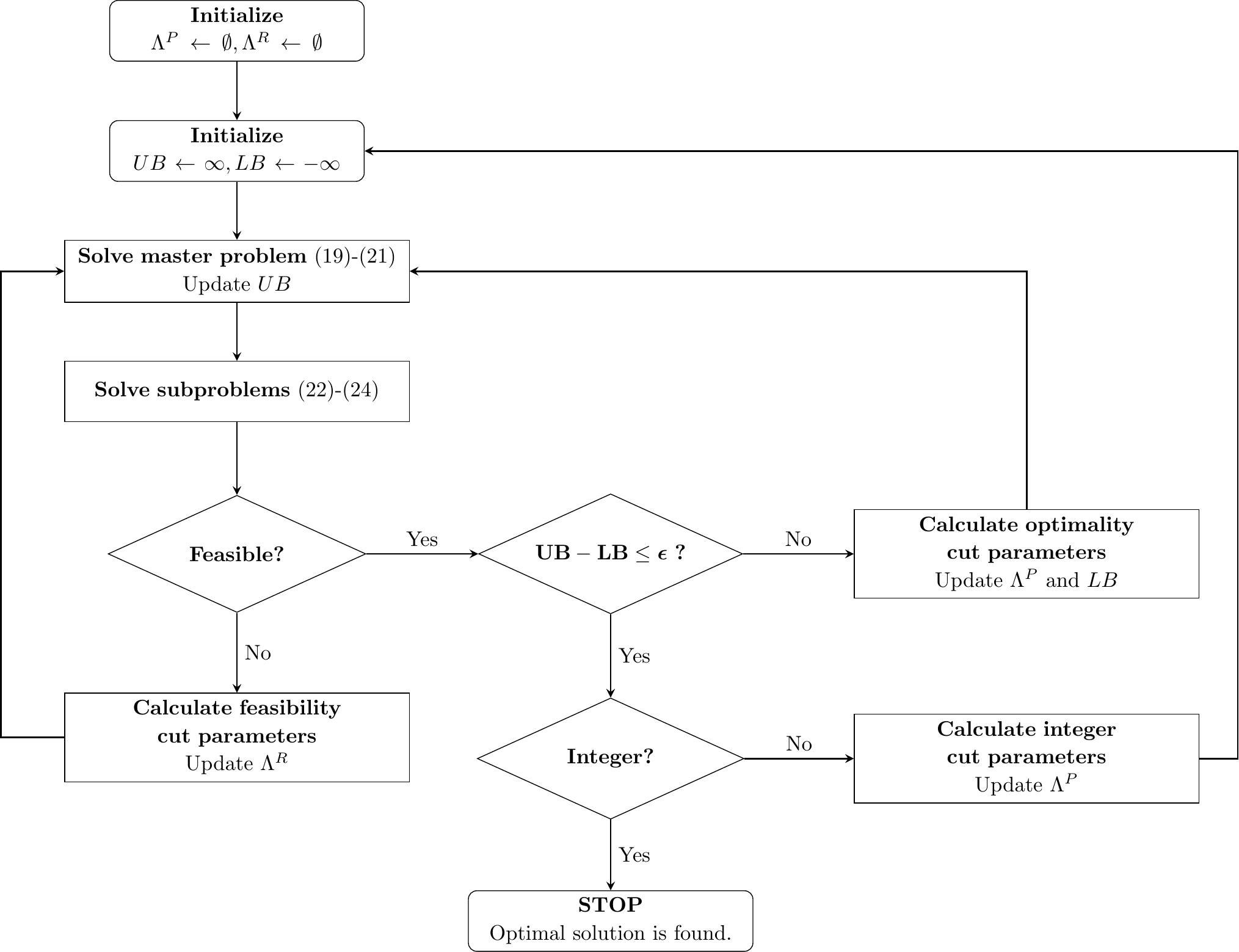}
\caption{Flow diagram for Algorithm \ref{solution_algorithm}}
\label{fig:benders}
\end{figure}

If the described procedure of iteratively solving master and subproblems terminates with integer second stage variables, the algorithm stops since it reaches the optimal solution. If one or more second stage variables have non-integer value, then an integer cut of type \cite{laporte1993integer} is added to penalize non-integer values in the given solution. Without loss of generality, this cut is incorporated within $\Psi \geq \boldsymbol{\lambda}^T \boldsymbol{\alpha}$, and the cut coefficient vector defining this integer cut is added to the set $\Lambda^P$. The steps of the described solution algorithm is summarized in Algorithm \ref{solution_algorithm}, and a flow diagram is given in Figure \ref{fig:benders}.

The proposed solution algorithm converges to the optimal solution in a finite number of iterations. This is a straightforward result due to {\em(i)} finite convergence of Benders' decomposition with integer cuts, and {\em(ii)} the finite number of integer variables (denoted by $\boldsymbol{\alpha}$ that couple restricted master problem with the subproblem). This result is demonstrated using two key observations. First, for any feasible solution to $\boldsymbol{\alpha}$; one can obtain a Benders' cut that imposes the cost of the continuous relaxation of the subproblem, and an integer cut that imposes the exact cost of the subproblem. While imposing exact costs of the corresponding subproblems for a specific $\boldsymbol{\alpha}$, these cuts also generate a lower bound for any other feasible solution $\boldsymbol{\alpha}'$. Second, due to the finite number of integer variables, one needs to execute a finite number of iterations before convergence is achieved. More specifically, if there are $N$ many $\boldsymbol{\alpha}$ variables, then a total number of $2^N$ Benders' and $2^N$ integer cuts would be sufficient in worst case scenario to recover the entire cost of the subproblems. For a detailed proof of finite convergence and optimality in a similar problem structure, the reader is referred to \cite{yildirim2}.

\begin{algorithm}
\caption{Solution Algorithm} \label{solution_algorithm}
\begin{algorithmic}
    \STATE $IsInteger := False$
    \WHILE{$IsInteger = False$}
    \STATE $LB := -\infty$, $UB := \infty$
        \WHILE{$UB - LB > \epsilon$}
            \STATE Solve master problem \eqref{master_obj}-\eqref{feas_cut}; obtain solution vector $(\bar{x}, \bar{y}, \bar{\alpha})$ and objective value
            \STATE Solve relaxed and MIP forms of the subproblem \eqref{sub_obj}-\eqref{sub_C8} using $\bar{\alpha}$ values
            \IF{subproblem returns a feasible solution}
            \STATE Update $UB$ by summing the current master and MIP subproblem objective values
                \STATE Obtain dual vector $\lambda$; add to $\Lambda^P$
                \STATE Update $LB$ based on relaxed subproblem objective value
            \ELSIF{subproblem has no feasible solutions}
                \STATE Dual of the subproblem is unbounded; obtain dual extreme ray $\lambda$, add to $\Lambda^R$ 
            \ENDIF
        \ENDWHILE
        \IF{solution is integer}
            \STATE $IsInteger = True$
        \ELSE
            \STATE Compute integer cut coefficients, add to $\Lambda^P$.
        \ENDIF
    \ENDWHILE
\end{algorithmic}
\end{algorithm}

To test the performance of the decomposition method, representative experiments for different cases are conducted. Table \ref{table:benders} shows analyses under different number of wind farms, wind turbines, and turbine components. All computational figures are average of $5$ replication runs. In all considered instances, decomposition proves faster than the standard model. It is also noteworthy to mention that the decomposition method is able to provide solutions in reasonable times for instances where no feasible solution is found by the standard model. This structural decomposition is especially important with the increasing number of turbines and wind farms to be operated. It is concluded that the proposed solution algorithm is beneficial and experiments can be continued by this technique.

\begin{table}
	\centering
	\caption{Computational Performance: Full Model vs. Solution Algorithm for Problems with 4- and 8-Component Turbines} \label{table:benders}
	\renewcommand{\arraystretch}{1.2}
	\begin{tabular}{|C{30pt}|C{50pt}|C{80pt}|C{80pt}|}
		\hline $\boldsymbol{|\mathcal{K}|}$ & $\boldsymbol{|\mathcal{L}|, |\mathcal{G}|}$ & {\bf Full Model} & {\bf Solution Algorithm} \\ \hline \hline
		\multirow{4}{*}{4} & 2, 150 & 199.01 & 49.50 \\ \cline{2-4}
		& 3, 100 & 226.97 & 34.34 \\ \cline{2-4}
		& 4, 100 & 687.55 & 453.88 \\ \cline{2-4}
		& 3, 150 & 957.81$^{*}$ & 771.02 \\ \hline
		\multirow{4}{*}{8} & 2, 150 & 628.76 & 179.04 \\ \cline{2-4}
		& 3, 100 & 589.09 & 132.18 \\ \cline{2-4}
		& 4, 100 & 1311.85 & 819.13 \\ \cline{2-4}
		& 3, 150 & 1769.60 & 689.42 \\ \hline
		\multicolumn{4}{l}{\footnotesize{\shortstack[l]{Results indicate average computation time \emph{(seconds)} across 5 replications. \\ $^{*}$In 2 of the 5 replications, a feasible solution could not be found within \\ \hspace{2pt} the specified time limit of one hour.}}}
	\end{tabular}
\end{table}

\section{Experimental Results} \label{sec:experiments}

In this section experiments are performed to study the performance of different maintenance polices in a wide range of settings and conditions. The data used in the experiments can be introduced as follows: \textit{Degradation Data -} Rotating machinery degradation database from \cite{gebraeel2005residual} is used to emulate the degradation process in the wind turbine systems. Rotating machinery are run from brand new to failure, and their raw vibration spectra are captured continuously. These raw signals are then translated into degradation signals that are used to update the predictions on component remaining life. For more information on the specifics of the database, we refer the reader to \cite{gebraeel2005residual}. Rotating machinery degradation (e.g. bearing degradation) is an important contributor to turbine failure.  However, turbines have many other distinct components, each with a different degradation process and a unique set of sensors for monitoring them (e.g. electrical or vibration based) \cite{marquez2012condition,hameed2009condition,salameh2018gearbox}. The focus is to leverage on this existing literature, and build adaptive optimization models that can adapt to the resulting predictions on component remaining life, regardless of the underlying degradation processes and sensor requirements.  \textit{Weather Data -} Weather data from the KNMI North Sea Wind Atlas project \cite{KNMI2014} is used. This data set consists of hourly wind speed measurements at $8$ different height levels ($10$, $20$, $40$, $60$, $80$, $100$, $150$ and $200$ m) over a time period of January 2014 up to January 2017. The hourly wind speed data is adapted for the presented maintenance optimization model, where each time period corresponds to two days, by computing the average wind speed of each consecutive 2-day period. Measurements and parameters are based on an existing wind farm, GEMINI, in the North Sea, consisting of 150 Siemens SWT 4.0-130 turbines with a generating capacity of 4 MW and a rotor diameter of 130 meters. The power output is estimated from the KNMI wind speed measurements using the wind turbine power curve model \cite{Lydia2014} of SWT 4.0-130 turbines (with cut-in, rated, and cut-off wind speeds of 5, 12, and 25 m/s, respectively). 

\begin{figure}[t!]
\centering
\includegraphics[width=0.85\textwidth]{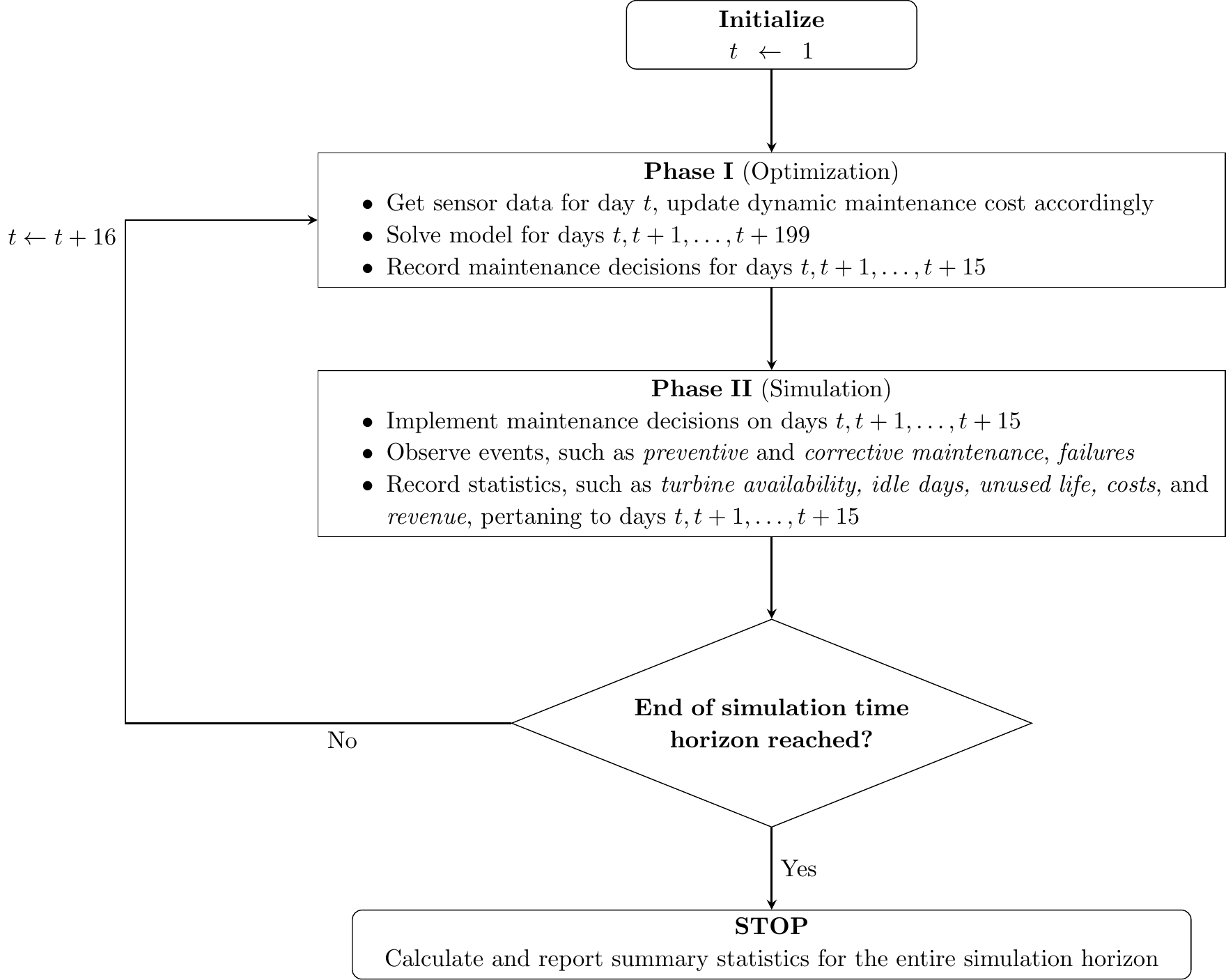}
\caption{Flow diagram of the rolling horizon framework}
\label{fig:rolling_horizon}
\end{figure}

The rolling horizon based experimental framework, which is summarized in Figure \ref{fig:rolling_horizon}, is formed by two phases. In the first phase the  optimization model (given in Section \ref{sec:methodology}) is solved with the tailored solution method (given in Section \ref{sec:algorithm}) to schedule the operations and maintenance actions over a 200-day planning horizon, denoted as 100 equal time periods of 2 days. Here, the dynamic maintenance costs of operational turbine components are used. In the second phase, the chain of events are simulated based on the rotating machinery degradation data for a sequential fixed number of time periods (16 days). For every day of this fixed period, the status of wind turbines is observed and recorded: whether they undergo preventive or corrective maintenance, whether a failure occurs, and whether the turbines have idle periods. For components that experience preventive maintenance, how long the component had before failure is calculated -- a term referred to as {\em unused life}. An unexpected failure occurs when an operational component was scheduled for a preventive maintenance but fails (i.e., is degraded beyond its failure threshold) before the scheduled preventive maintenance could take place. The wind turbines that have at least one failed component, and thus was in failed state at the beginning of the planning horizon, remain idle until their failed components undergo corrective maintenance. For each time period of the second phase, the metrics for revenues, costs, detailed maintenance actions, and statistics such as idle days, availability of the wind turbines, failures, and unused life are recorded. At the end of the second phase, dynamic maintenance costs for each component in operational wind turbines are updated with the most recent sensor readings.

\subsection{Comparative Performance Evaluation of MC-CBOM} \label{sec:exp:mc-cbom}

To demonstrate the effectiveness of the approach, a series of comparative studies that evaluate the performance of the proposed model vs. benchmarks with different properties is conducted. In the experiments, a large scale system composed of $3$ wind farms with $100$ turbines is considered. Each turbine is composed of $4$ components, amounting to $1200$ components in total. First two sets of benchmarks (given in Sections \ref{sc-cbm} and \ref{bc-cbm}) refer to sensor-driven approaches that have equal predictive power to the proposed model in terms of predicting failure likelihoods. The final set of benchmarks (given in Sections \ref{pm} and \ref{rm}) refer to more traditional maintenance policies that do not adapt to any sensor information.
\subsubsection{Single Component - Condition Based Maintenance (SC-CBM$_1$ \& SC-CBM$_2$)} \label{sc-cbm} 
These models maintain a single component per turbine visit. This is representative of a sensor-driven model that does not use opportunistic cost across turbine components.  
In the first variant SC-CBM$_1$, the following constraints are used to ensure that at most a single component is maintained at a time:
\begin{equation}
 \sum_{k \in \mathcal{K}_o^{l,i}} z_t^{l,i,k}+\sum_{k \in \mathcal{K}_f^{l,i}} \nu_t^{l,i,k} \leq 1, \; \forall l \in \mathcal{L}, \forall i \in \mathcal{G}^l, \forall t \in \mathcal{T} \label{cbom-1}
\end{equation} 
whereas in the second variant SC-CBM$_2$, this limit is restricted only to preventive maintenances (i.e. allowing single preventive and multiple corrective maintenances simultaneously). This variant requires suppressing the second term in \eqref{cbom-1}.
This setting is most appropriate for cases where the turbine visit cost is negligible compared to the costs of component maintenance.

\subsubsection{Batched Component - Condition Based Maintenance (BC-CBOM$_1$ \& BC-CBOM$_2$)} \label{bc-cbm}
In these models, maintenances on all operational components must be conducted in a single visit to a turbine. BC-CBOM$_1$ enforces this rule to all the components, whereas BC-CBOM$_2$ enforces batching only for the components that are preventively maintained. For BC-CBOM$_2$, the following constraints are added:

\vspace{-20pt}
\begin{equation}
z_t^{l,i,k} = z_t^{l,i,k'}, \quad \forall l \in \mathcal{L}, \forall i \in \mathcal{G}^l, \forall k, k' \in \mathcal{K}^{l,i}_o, \forall t \in \mathcal{T} \label{cbom-all}
\end{equation}
\vspace{-20pt}

Likewise, the other variant BC-CBOM$_1$ enforces additional constraints using $\nu_t^{l,i,k}$ for components in $\mathcal{K}^{l,i}_f$. This setting is important when turbine visit cost is significant and should be minimized through aggressive batching.

\subsubsection{Periodic Maintenance (PM)} \label{pm} 
The PM model ensures that a component goes through preventive maintenance when its age is within a range that is optimized using the component failure times.

To implement this policy, constraints \eqref{C5} is replaced with ones that enforce PM limits, instead of dynamic limits, $\zeta^{l,i,k}$. 

PM model does not use any insights from sensor-driven dynamic maintenance cost functions. Therefore,  $C_{t^o_k,t}^{l,i,k} = 0, \quad \forall l \in \mathcal{L}, \forall i \in \mathcal{G}^l, \forall k \in \mathcal{K}^{l,i}$. This policy is useful for understanding the conventional practice in wind farm maintenance.

\subsubsection{Reactive Maintenance (RM)} \label{rm} 
The RM model does not conduct any preventive maintenance actions. The rest of the model is identical to the proposed model. To implement this policy,  constraints \eqref{C5} are removed from the MC-CBOM and $z_t^{l,i,k} = 0, \quad \forall l \in \mathcal{L}, \forall i \in \mathcal{G}^l, \forall k \in \mathcal{K}^{l,i}, \forall t \in \mathcal{T}$ is added.

\begin{figure}
\centering
\includegraphics[width=0.75\textwidth]{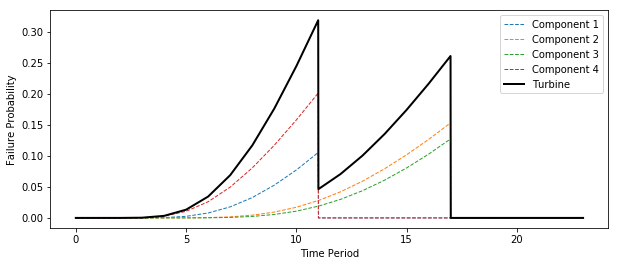}
\caption{Component and Turbine Failure Probabilities}
\label{fig:failure_prob}
\end{figure}

A key advantage of the proposed MC-CBOM model over the conventional models is the explicit modeling of the relationship between i) component outage risks, ii) turbine maintenance visit schedule, iii) component maintenance decisions, and iv) turbine failure risks. Figure \ref{fig:failure_prob} illustrates the dynamic progression of component and turbine failure probabilities in one of the runs as a function of maintenance decisions. For this particular scenario, turbine $i$ at wind farm $l$ has maintenance visits scheduled at time periods $11$ and $17$ (i.e. $\alpha_{11}^{l,i} = \alpha_{17}^{l,i} = 1$); maintaining components $1$ \& $4$ in the first visit ($z_{11}^{l,i,1} = z_{11}^{l,i,4} = 1$), and $2$ \& $3$ in the second visit ($z_{17}^{l,i,2} = z_{17}^{l,i,3} = 1$). Note that both turbine and component probabilities change as a function of turbine visit and component maintenance decisions. The costs associated with these maintenance decisions, in turn, are continuously updated by sensor information.

\subsubsection{Case Results}

\begin{table}[!htb]
	\caption{Performance of MC-CBOM Compared to Other Sensor-Based Maintenance Policies} \label{table:benchmark}
	\renewcommand{\arraystretch}{1.2}
		\begin{tabular}{|l | c | c c c c|}
		    \cline{2-6} \multicolumn{1}{c|}{} & & \multicolumn{4}{c|}{\textbf{Other Sensor-Based Models}} \\
		    \multicolumn{1}{c|}{} & \multicolumn{1}{c|}{\textbf{MC-CBOM}} & \textbf{SC-CBM$_1$} & \textbf{SC-CBM$_2$} & \textbf{BC-CBOM$_1$} & \textbf{BC-CBOM$_2$} \\
			 \hline
			Net Profit & \$63.05M & \$61.15M & \$61.19M & \$56.90M & \$57.54M \\ \hline
			Revenue & \$100.40M & \$100.00M & \$100.02M & \$101.77M & \$102.19M \\ \hline
			Expenditures & \$37.35M & \$38.85M & \$38.82M & \$44.87M & \$44.66M  \\ \hline
			\hspace{5pt} - Component Maint. & \$25.13M & \$25.16M & \$25.14M & \$38.50M & \$38.23M  \\ \hline
			\hspace{5pt} - Location Visit & \$2.71M & \$2.87M & \$2.88M& \$2.19M & \$2.18M \\ \hline
			\hspace{5pt} - Turbine Visit & \$9.51M& \$10.83M& \$10.80M& \$4.18M& \$4.24M \\ \hline \hline
			\# Preventive Actions & 2990 & 2990 & 2982 & 4716 & 4693  \\ \hline
			\# Component Failures & 101 & 103 & 107 & 64 & 57  \\ \hline
			Unused Life & 9.73 & 9.76 & 9.75 & 51.12 & 50.61  \\ \hline
			Avg. Batch Size & 1.138 & 1.000 & 1.001 & 4.000 & 3.919 \\ \hline
	\end{tabular}
\end{table} \vspace{10pt}

\begin{table}[!htb]
    \caption{Performance of MC-CBOM Compared to Reliability-Based Maintenance Policies} \label{table:benchmark2}
	\renewcommand{\arraystretch}{1.2}
		\begin{tabular}{|l | c | C{2cm} C{2cm} |}
            \cline{2-4} \multicolumn{1}{c|}{} & & \multicolumn{2}{c|}{ \textbf{Reliability-Based Models}} \\
		    \multicolumn{1}{c|}{} & \multicolumn{1}{c|}{\textbf{MC-CBOM}} & \textbf{RM} & \textbf{PM} \\ \hline
			Net Profit & \$63.05M & \$45.26M & \$39.41M \\ \hline
			Revenue & \$100.40M & \$86.22M & \$97.14M \\ \hline
			Expenditures & \$37.35M & \$40.97M & \$57.73M  \\ \hline
			\hspace{5pt} - Component Maint. & \$25.13M & \$31.40M & \$45.24M  \\ \hline
			\hspace{5pt} - Location Visit & \$2.71M & \$1.23M & \$1.09M \\ \hline
			\hspace{5pt} - Turbine Visit & \$9.51M & \$8.33M& \$11.40M \\ \hline \hline
			\# Preventive Actions & 2990 & 0 & 4871  \\ \hline
			\# Component Failures & 101 & 2617 & 523  \\ \hline
			Unused Life & 9.73 & 0 & 67.24  \\ \hline
			Avg. Batch Size & 1.138 & 1.099 & 1.656  \\ \hline
	\end{tabular}
\end{table}

Tables \ref{table:benchmark} and \ref{table:benchmark2} provide results for different maintenance policies. Note that the MC-CBOM provides significant benefits in terms of cost and reliability. MC-CBOM's profit is $\geq 3.03\%$ better than SC-CBM policies, and $\geq 9.58\%$ better than BC-CBOM policies. This is accomplished through significant savings in maintenance costs (by $\geq 4.30\%$ and $\geq 16.8$ for SC-CBM and BC-CBOM, respectively), without compromising operational revenue. As expected, SC-CBM and BC-CBOM policies maintained approximately $1$ and $4$ components per turbine visit, respectively. Average number of visits in these policies are enforced through stylized rules. In reality, the average number of components replaced per turbine visit is a complex function of operational requirements, turbine visit costs, and the opportunity cost associated with market revenue. The proposed MC-CBOM model, adapts to these cost factors to identify a tradeoff point at $1.138$ components per turbine. 

The comparison of MC-CBOM with SC-CBM and BC-CBOM models indicate that access to degradation models and accurate prediction on component and turbine failure likelihoods does not necessarily provide improvements in fleet-level maintenance. While these predictions are important, their full potential can be recovered only if the complex interactions between different components, turbines and wind farm locations, are explicitly considered with the operations and maintenance optimization model.

Note that MC-CBOM performs significantly better compared to the conventional (i.e. reliability-based) PM and RM models that do not use sensor information. MC-CBOM provides significant improvements in profit, and reduces maintenance cost by $8.83\%$ and $35.30\%$, respectively. Note that both RM and PM policies perform poorly without access to the sensor-driven insights on component failure likelihoods. Lack of these insights lead to unnecessary and early maintenances (for PM) and increased number of failures (for RM and PM).

\subsection{Impact of Electricity Price} \label{sec:exp:electricity_price}

The next set of experiments demonstrate the impact of market price on the performance of MC-CBOM. Table \ref{table:cbom} presents the results for different price scenarios. Note that as market price increases, MC-CBOM increases the average batch size (i.e., the average number of components maintained during a turbine visit) to limit the downtime due to maintenance -- by $5.88\%$ between the scenarios with market prices of $12.5$ \$/MWh and $50$ \$/MWh. This results in an increase in availability and unused life -- by $0.15\%$ and $5.23\%$ between $12.5$ \$/MWh and $50$ \$/MWh, respectively. These changes also impact component maintenance cost, which exhibits a small increase of $0.76\%$ between $12.5$ \$/MWh and $50$ \$/MWh. This experiment shows that as turbine availability becomes more significant, MC-CBOM deviates slightly from its optimal maintenance policy to minimize turbine outages.
\begin{table}[H]
	\centering
	\caption{Impact of Electricity Price on MC-CBOM} \label{table:cbom}
	{
		\begin{tabular}{|l | c c c c|}
			\hline \multicolumn{1}{ |l| }{ \textbf{Market Price (\$/MWh) }}
			 & \textbf{12.5} & \textbf{25} & \textbf{37.5} & \textbf{50} \\ \hline
			Component Maint. Costs & \$25.06M & \$25.14M & \$25.19M & \$25.25M  \\ \hline
			Availability & 96.86\% & 96.86\% & 96.96\% & 97.00\% \\ \hline
			Unused Life & 9.75 & 9.96 & 10.02 & 10.26  \\ \hline
			Avg. Batch Size & 1.140 & 1.162 & 1.184 & 1.207 \\ \hline
	\end{tabular}}
\end{table}

\subsection{Impact of Component Significance} \label{sec:exp:component}

In the last set of experiments, presented in Table \ref{table:cbom_comp}, how the relative cost of different components impact their respective maintenance schedules is studied. Components C1 through C4 denote the gearbox components, rotor, generator, and bearing, with associated preventive maintenance costs of \$$38k$, \$$28k$, \$$25k$, and \$$15k$ (as in \cite{Ding2012}).

The key observation of this section is that while all components deviate from their component-specific optimal maintenance schedules, components that have less significance would be more liberal in adapting to turbine- and fleet-level requirements (deviations increase by $76\%$ between C1 and C4). It is also observed that low cost components would initiate less turbine-level maintenances (by $27\%$ between C1 and C4), and appear less often within turbine maintenances with single component actions (by $13\%$ between C1 and C4). This experiment demonstrates that the proposed model can effectively incorporate how the importance of components play a significant role in determining the degree of their interaction with turbine- and fleet-level decisions.
\begin{table}[H]
	\centering
	\caption{Impact of Component Significance on MC-CBOM} \label{table:cbom_comp}
	{
		\begin{tabular}{|l | c c c c|}
			\hline \multicolumn{1}{ |l| }{ \textbf{{Component \#}
			}} & \textbf{C1} & \textbf{C2} & \textbf{C3} & \textbf{C4} \\ \hline
			Average Deviation from Optimal CBM & 4.66 & 5.92 & 6.12 & 8.21  \\\hline
			\# of Maint. Initiated by Component & 614 & 575 & 566 & 511  \\\hline
			\# of Maint. by a Single Component & 356 & 350 & 341 & 309  \\ \hline
	\end{tabular}}
\end{table}

\subsection{Discussion of Findings}

An in-depth comparative performance analysis of the proposed MC-CBOM policy, as well as insights into how it adapts to varying conditions, is provided in Sections \ref{sec:exp:mc-cbom}, \ref{sec:exp:electricity_price}, and \ref{sec:exp:component}. The MC-CBOM policy outperforms all considered benchmark policies thanks to its ability to consider and adapt to the complex interactions between different decision layers. By making use of real-time sensor information, it is able to significantly outperform traditional RM and PM policies, since these tend to either conduct maintenance too late, which results in low availability and production, or too early, which results in high maintenance costs. Additionally, it is observed that the flexibility of MC-CBOM in identifying which components/turbines are to be maintained during a crew visit, results in lower maintenance expenditures and higher profits compared to more restrictive condition-based policies, such as SC-CBM and BC-CBOM.

The MC-CBOM optimization model is adaptable to changing problem parameters, such as electricity price, maintenance costs, turbine/location visit costs, fleet/crew size, and weather conditions. As an example of how it adapts to changing conditions, Section \ref{sec:exp:electricity_price} provides the outcome of MC-CBOM under varying electricity prices. Table \ref{table:cbom} demonstrates that as electricity price increases, MC-CBOM adapts by pursuing a more aggressive maintenance strategy and therefore increasing overall availability of the turbines.

An important contribution of this study is to consider maintenance at a component, rather than at a turbine, level. This allows the MC-CBOM model to adapt to different component characteristics by distinguishing between turbine components based on their maintenance costs and degradation levels. This feature of the proposed method is quantitatively demonstrated in Section \ref{sec:exp:component}. It is observed that MC-CBOM prioritizes the components that are more "critical", i.e., more expensive to maintain, and schedules preventive maintenance on these critical components as close to their individual optimal maintenance time as possible. This consideration of component criticality results in lower maintenance expenditures than methodologies with only turbine-level maintenance schedules.

In order to gain further insight into the impact of the MC-CBOM model, interesting directions for future work would be i) a comprehensive computational study that focuses primarily on testing the performance of the proposed framework under a wide range of parameter values to highlight its practical implications for the diverse needs of the practitioner community, and ii) a variant of the MC-CBOM focusing on shorter timescales (e.g., hours), which captures the interactions between maintenance decisions and degradation due to short-term wind fluctuations.

\section{Conclusion} \label{sec:conclusion}
A unified and scalable framework for integrating multi-component degradation models to larger scale wind farm systems is proposed in this paper. In doing so, a novel optimization model, MC-CBOM, is developed, that adapts to real-time sensor information while accounting for complex dependencies across components, turbines and wind farms. To ensure scalable deployment of the proposed model, a solution algorithm that exploits the structure of the reformulated optimization model is devised to obtain the optimal O\&M policy in an efficient manner. 

A simulation framework based on the rolling horizon methodology is used to extensively test the performance of the proposed optimization model in terms of net profit, maintenance costs, number of failures, unused life, and turbine availability. The proposed simulation framework builds on a real-world degradation database for rotating machinery and weather data from the KNMI North Sea Wind Atlas project \cite{KNMI2014} to create realistic test cases for testing. It is quantitatively demonstrated that the proposed framework provides significant cost and reliability improvements over existing maintenance models. Furthermore, the ability of the proposed model to adapt to a wide range of operational and maintenance scenarios is illustrated. More specifically, it is observed that {\em(i)} the MC-CBOM model adapts to increasing electricity price by pursuing a more aggressive maintenance strategy in order to ensure high turbine availability, and {\em(ii)} the multi-component nature of MC-CBOM is effective in prioritizing critical components over others in order to reduce overall maintenance expenditure.

\section*{References}
\bibliography{main}

\end{document}